# Segmentation of Shoulder Muscle MRI Using a New Region and Edge based Deep Auto-Encoder

Saddam Hussain Khan[1, 2], Asifullah Khan[1, 2, 3*], Yeon Soo Lee[4], Mehdi Hassan[5], and Woong Kyo jeong[6].

asif@pieas.edu.pk

[1]Pattern Recognition Lab, Department of Computer & Information Sciences, Pakistan Institute of Engineering & Applied Sciences, Nilore, Islamabad 45650, Pakistan,

[2]PIEAS Artificial Intelligence Center (PAIC), Pakistan Institute of Engineering & Applied Sciences, Nilore, Islamabad 45650, Pakistan

[3]Center for Mathematical Sciences, Pakistan Institute of Engineering & Applied Sciences, Nilore, Islamabad 45650, Pakistan

[4]Department of Biomedical Engineering, College of Medical Science, Catholic University of Daegu, South Korea

Medical Teaching Institute, Peshawar, Pakistan

[5]Department of Computer Science, Air University, Islamabad, Pakistan

[6]Department of Orthopaedic Surgery, Korea University College of Medicine, Seoul, Korea.

## ABSTRACT

Automatic segmentation of shoulder muscle MRI is challenging due to the high variation in muscle size, shape, texture, and spatial position of tears. Manual segmentation of tear and muscle portion is hard, time-consuming, and subjective to pathological expertise. This work proposes a new Region and Edge-based Deep Auto-Encoder (RE-DAE) for shoulder muscle MRI segmentation. The proposed RE-DAE harmoniously employs average and max-pooling operation in the encoder and decoder blocks of the Convolutional Neural Network (CNN). Region-based segmentation incorporated in the Deep Auto-Encoder (DAE) encourages the network to extract smooth and homogenous regions. In contrast, edge-based segmentation tries to learn the boundary and anatomical information. These two concepts, systematically combined in a DAE, generate a discriminative and sparse hybrid feature space (exploiting both region homogeneity and boundaries). Moreover, the concept of static attention is exploited in the proposed RE-DAE that helps in effectively learning the tear region. The performances of the proposed MRI segmentation based DAE architectures have been tested using a 3D MRI shoulder muscle dataset using the hold-out cross-validation technique. The MRI data has been collected from the Korea University Anam Hospital, Seoul, South Korea. Experimental comparisons have been conducted by employing innovative custom-made and existing pre-trained CNN architectures both using transfer learning and fine-tuning. Objective evaluation on the muscle datasets using the proposed SA-RE-DAE showed a dice similarity of 85.58% and 87.07%, an accuracy of 81.57% and 95.58% for tear and muscle regions, respectively. The high visual quality and the objective result suggest that the proposed SA-RE-DAE is able to correctly segment tear and muscle regions in shoulder muscle MRI for better clinical decisions.

# 1. Introduction

Over the last two decades, there has been a significant advancement in medical imaging technology and pathological analyses [1]. Medical image processing (MIP) plays a vital role in the diagnosis and screening of diseases. In the anatomical imaging community, pixel-based segmentation of medical images generally classifies pixels into different anatomic regions [2]. These anatomic regions are bones, muscles, and blood vessels, etc. Applications of segmentation include effective diagnosis, surgery, vision boosting, and the extraction of qualitative and quantitative measures from medical datasets [3], [4]. However, the manual study of magnetic resonance imaging (MRI) based images created under different contrast or intensity can be quite tricky and time-consuming for radiologists and researchers. Consequently, computer-aided diagnostics exploit automatic image segmentation techniques, providing a guideline for image interpretation and making a quick and accurate decision [5], [6].

Nowadays, many people suffer from rotator cuff tear problems, which cause severe pain and limit normal activities like motion and strength of muscles. There are four muscles named infraspinatus, subscapularis, supraspinatus, and teres minor, which comprise the rotator cuff and cover the shoulder's inner space [7]. Injuries of muscles occur gradually due to continued computer usage, sports like chin up, cliff climbing, swimming, and some degenerative shoulder diseases [8]. In this context, consequently, the numbers of rotator cuff tear patients are increasing day by day. Therefore, efficient diagnosis and surgery of rotator cuff have become essential. However, the lack of muscle recovery after the operation brings about severe and complicated problems [9]. Simultaneously, accurate quantitative measurement methods of an anatomical structure are crucial to assess the pre- and post-operation phases of the issues. Previous studies proposed several classical and automatic methods [10], [11]. A classical active contour segmentation approach has been employed to extract the supraspinatus muscle configuration [10]. The reported work achieved a dice similarity (DS Score) of (95%) for supraspinatus muscle using the 3D magnetic resonance (MR) muscle dataset [12]. However, the active contour method required user input by initializing region (supraspinatus muscle) and parametric values. Moreover, automated segmentation of shoulder MR images using deep learning (DL) techniques has been scarcely applied to supraspinatus muscle [13].

The current study mainly focuses on the exploitation of the learning capabilities of DL approaches for the shoulder muscle segmentation of MR images [5], [14], [15]. The correct segmentation of MR images is always challenging due to a sudden change in intensity, contrast level, size, shape, and global MR appearance of the physical structure of a muscle [16]. The main difficulty is in segmenting regions with missing edges and absence of texture contrast as well in separating the region of interest (ROI) from muscles and background. To address such segmentation complexities, most of the earlier systems were built on traditional segmentation methods [17]. Generally, these traditional segmentation approaches were based on edge detection or mathematical modeling and quite effective under normal circumstances [18]–[21]. However, these methods have some limitations in generating correct segmentation due to intensity inhomogeneity, higher noise level, and uneven illumination. The features-based machine learning (ML) approaches were introduced to address these challenges since these showed improved segmentation.

Additionally, automatic supervised and unsupervised segmentation based on ML approaches were reported to extract hand-crafted features [22]–[29]. Designing and extracting these hand-crafted features have always been the primary concern for developing a correct segmentation system. However, hand-crafted features were tested mainly on a small dataset. The complexities of these approaches have been considered as a significant limitation for them to be deployed. On the other hand, DL achieved high visual quality results compared to traditional ML techniques in the field of MIP, especially in medical MRI segmentation [30]–[33]. Recently, DL has been employed for rotator cuff shoulder muscles (subscapularis, supraspinatus, infraspinatus, deltoid) segmentation [11]. The modified U-Net (VGG-16's encoder and U-Net's decoder) has been employed on the OBPP muscle dataset [34]. The modified U-Net achieved DS scores of 82.4%, 82.0%, 71.0% and 82.8% for deltoid, infraspinatus, supraspinatus and subscapularis muscles, respectively. Importantly, tears of supraspinatus tendons may also occur, and degeneration of muscles limits normal activities. However, these methods have some limitations in providing accurate tear-related measurements associated with rotator cuff volume and thus achieve a better tear problem. Furthermore, the aforementioned methods addressed rotator cuff muscles and supraspinatus muscle segmentation but not addressed the tear-related challenges.

In this research work, we have developed a fully automated muscle segmentation system, named "Region and Edge-based Deep Auto-Encoder (RE-DAE)", based on Convolutional Neural Network (CNN). The proposed RE-DAE is developed and evaluated on the tear-related shoulder muscle dataset. The significant contributions of this research study are as follow.

1. To the best of our knowledge, this is the first study that incorporates region and edge-based pixel-wise segmentation in CNN architectures. Thus, it proposes a fully automatic (RE-DAE) tear and supraspinatus muscle segmentation architecture. The proposed RE-DAE systematically employs region and edge-based segmentation in each encoder and decoder block.
2. The static attention concept is introduced in the proposed RE-DAE to effectively learn the sparse representation of the tear region for better segmentation.
3. Custom-made CNN models based on the concept of semantic segmentation are developed and are trained from scratch.
4. To improve the convergence of the custom-made CNN models, the idea of transfer learning (TL) is employed during the segmentation. Fine-tuning of the custom CNN architectures is performed by borrowing weights from the corresponding layers of the existing pre-trained models.

The paper planned as follows. Section 2 discusses the detail of the proposed semantic segmentation framework. Section 3 presents an experimental setup. Section 4 describes the results, discussion, and comparative analysis. Finally, section 5 concludes the paper.

## 2. The Proposed Semantic Segmentation Framework

The proposed muscle segmentation framework aims to accurately and precisely segment the tear portion from the entire region of the muscle and background region. The workflow of the segmentation framework is divided into four main phases: (1) the pre-processing, which prepares the data necessary for training the segmentation networks, (2) the semantic segmentation models training phase, (3) the model deployment phase is evaluated by pixel-label-based semantic segmentation, and (4) the pixel attention to the proposed segmentation model. The aforementioned phases of the proposed segmentation framework are shown in Fig .1. The details of each phase are provided in the forthcoming sections.

## 2.1 Pre-processing

Pre-processing is essential in subsequent segmentation tasks, especially for medical image segmentation. Muscle MRI images predominantly suffer due to contrast or intensity variation, low illumination, clutter, and noise. Some pre-processing methods were employed to enhance the subjective quality of images and make the model training smooth and quantifiable before providing the segmentation network. During the pre-processing, three main preparatory steps were employed: (1) 3D to 2D slice conversion and dataset splitting, (2) contrast enhancement of the images by using Histogram equalization to improve the visual quality, and (3) and data augmentation of the original dataset.

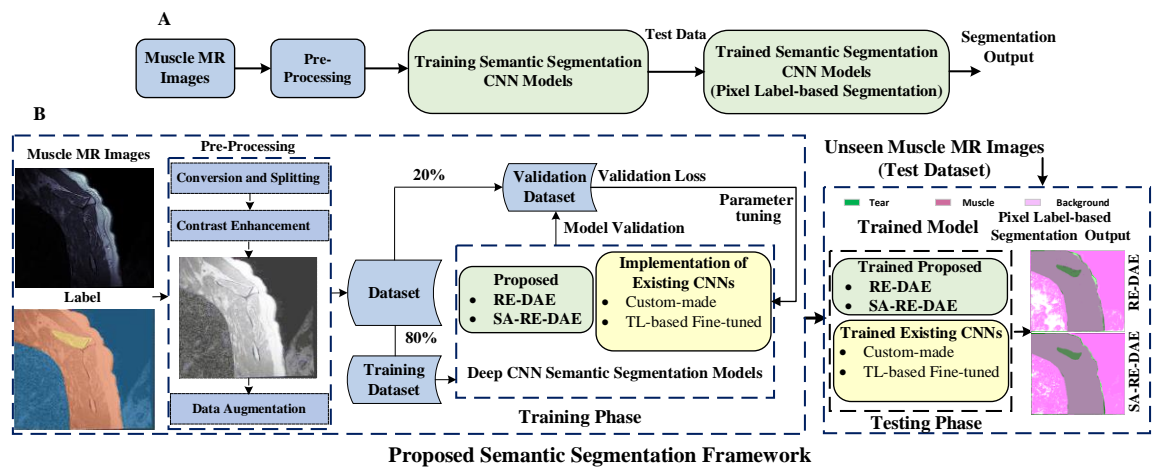

**Fig. 1.** (A) & (B) shows the brief and detailed (B) flow diagram of the proposed training and testing segmentation framework. The training images are preprocessed and then feed into the CNN network and their respective ground label.

### 2.1.1 Conversion and Splitting

The MRI images are in 3D format; therefore, the MRIcon software was used to convert 3D to 2D slices. Each slice represents a single MRI RGB image of size 304×304×3 pixels. The dataset was partitioned into training and testing at the ratio of 8:2. The training images and their corresponding label were used for training end to end CNN networks.

### 2.1.2 Contrast Enhancement

Histogram equalization of digital images stretches out the intensity range of an image into a broad range of intensity. Therefore, Histogram equalization has been applied in the pre-processing phase for enhancing the contrast and visual quality of an image, as shown in Fig .1.

### 2.1.3 Data Augmentation

There is a considerable chance of over-fitting when training a DL network on a small amount of dataset during the training phase. Therefore, data augmentation strategies were employed to increase the number of data samples. During training, the original and augmented images were combined and fed into CNN. While in on-the-fly strategy, different transformations like rotation, translation, scaling, and flipping was employed at run-time during training the CNNs. This type of data augmentation also helps in making the learner a robust one. The standard data augmentation, like a random reflection on both the X- and Y-axis, scaled factor at the range of [0.5 1], flipped along both the dimensions and rotated by ±10 degrees.

## 2.2 Training of Deep CNN Segmentation Networks

The deep semantic segmentation has three different segmentation experimental setups. (1) The proposed region and edge-based deep auto-encoder (RE-DAE), (2) Custom-made, and (3) TL-based fine-tuned deep CNN segmentation models.

### 2.2.1 The Proposed Region and Edge-based Deep Auto-Encoder (RE-DAE)

In semantic segmentation, every image is partitioned using pixel-wise labeling, where each pixel belongs to a specific class. Generally, the semantic segmentation approaches are used to understand indoor/outdoor scenes with many classes. However, these semantic segmentation architectures also play a key role in medical MR images with a few classes.

The proposed RE-DAE introduces the idea of how to exploit the region and edge-based pixel-wise semantic segmentation in CNN architecture, as shown in Fig. 2. In CNN networks, Convolutional layers employ trainable filters on the input images and create feature maps that help in learning the input contents as shown in eq. (1). The output feature maps are very sensitive to the position of the input contents. The small change in the position of the features in the input image will produce a different output. This problem might happen with shifting, scaling, flipping, and rotation of the original image. Therefore, the sub-sampling of feature maps overcomes the aforementioned problems. Sub-sampling is performed to achieve translation invariance for robust classification and the data having a small shifted variation. It is crucial to store spatial indices information while performing sub-sampling on the encoder side to achieve robustness. Based on stored indices, sub-sampling is performed in two different ways: max-pooling and average-pooling. When there is sufficient memory resource, then all the encoded features can be stored after sub-sampling.

**Edge and Region-Based Segmentation**

The edge-based operation allows a network to learn the image structure and boundaries information like edges, and textures, making discriminative sparse features set [35]. Max-pooling is employed as an edge operation that extracts boundaries of spatial information for practical application as well as also stores the location indices of high discriminative features as shown in eq. (2). The location indices are memorized in each pooling window at every encoder feature map. The pooling operation works independently on every input feature map and reduces the spatial dimension. When the stride size 'n' (non-overlapping window) is executed during the pooling operation, then the resultant feature map is down-sampled by a factor of 'n' along both width and height, thus reducing $\frac{3}{4}$n% of the activations. Conversely, up-sampling is performed by using the un-pooling layer in the decoder network. Specifically, up-sampling is performed using the memorized location indices calculated during the max-pooling operation of the corresponding encoder as illustrated in eq. (3). This process removes the need for learning during up-sampling. The up-sampled feature maps are sparse, and when convolved with trainable filters, they will generate dense channels similar in size to the corresponding input resolution. Max-pooling is more efficient to store location indices as compared with memorizing the feature maps. Max-pooling uses less memory, which makes it suitable for practical applications. Region-based operation encourages the network to identify an object's smooth and regional homogeneity information by employing average-pooling [36], as shown in eq. (4). Smooth operator smoothens the region variations by computing the average values of each local region and suppresses the noise distortions acquired during MR image acquisition.

$$O_{a,b} = \sum_{i,j}^{p,q} f_{p,q} C_{a+p,b+q} \qquad (1)$$

$$\mathbf{O}_{a,b}^{max-Down} = f_{max-Down}(\mathbf{O}_{a,b}) \qquad (2)$$

$$\mathbf{O}_{a,b}^{max-Up} = f_{max-Up}(\mathbf{O}_{a,b}) \qquad (3)$$

$$\mathbf{O}_{a,b}^{avg} = f_{avg}(\mathbf{O}_{a,b}) \qquad (4)$$

In eq. (1), $\mathbf{C}$ is an input channel of size $(A \times B)$ and filter is represented by $\mathbf{f}$. The receptive field of the input channel is represented by $(a, b)$ with respect to $\mathbf{f}$, and $(p, q)$ shows the spatial dimension of the filter. Similarly $f_{max-Down}(.)$, $f_{max-Up}(.)$, and $f_{avg}(.)$ are max-pooling, un-sampling, and average operations, respectively. $O_{a,b}^{max-Down}$, $O_{a,b}^{max-Up}$, and $O_{a,b}^{avg}$ show their respective outputs.

**The Systematic operation of both Region and Edge-Based**

Max-pooling extracts salient and discriminative sparse features for object segmentation and classification. In comparison, the combined operations of both the max and average-pooling extract hybrid feature space (both discriminative as well as smoothing features) to improve the segmentation performance. On the other hand, individual operations of both max and average-pooling have their drawbacks. Max-pooling extracts the maximum value within the pooling region, which leads to an unacceptable result because sometimes, most of the pooling region elements are of high intensity. Then, selecting a single high-intensity value within the pooling region may vanish the distinguishable features after max-pooling.

Conversely, average-pooling operates better on a cluster mode with a single centroid. But, its performance may be affected by data distribution that contains more than one centroid or outliers. Also, in a case where the features set have equivalent positive and negative values, the resultant activation could be small and may not provide a significant advantage. Sometimes, variance in features is not essential, and incorporating both types of pooling will produce approximately the same results. Typically, images in nature are ever-changing, and there is a high possibility that the drawback of both the pooling operations employed individually may have adverse effects on CNNs.

Therefore, in this work, the combined systematic operations of average local pooling (region) and max-pooling (edge) concepts are employed in the proposed RE-DAE as illustrated in eq. (5). Normally, max-pooling skip the regional information, which is recovered by average-pooling. Similarly, the basic intuition behind max-pooling is that with the loss of edge features due to the average operation, the network recovers all the edge information of an object. The excellent addressing ability is that the proposed combination of the pooling layers controls the number of features to learn the CNN network and avoids overfitting [35]. Experimental results demonstrate that the proposed combination of region and edge-based operation yields information-rich feature space, superior to individually applying the region and edge-based operations. The proposed combination has two main steps: (1) preserves smooth or regional information, (2) extracts high-intensity edge and discriminative features as depicts in eq. (5 & 6). When choosing pooling strategies, there is always a trade-off between information and noise. Since noise can be a clue for featuring, the noise will not be suppressed.

$$F_{RE-e} = f_c\left( \mathbf{O}_{x,y}^{avg} || \mathbf{O}_{x,y}^{max-Down} \right) \quad (5)$$

$$F_{RE-d} = f_c\left( \mathbf{O}_{x,y}^{max-Up} || \mathbf{O}_{x,y}^{avg} \right) \quad (6)$$

Finally, $\mathbf{F}_{RE-e}$ and $\mathbf{F}_{RE-d}$ are region and edge-based operation at encoder (e) and decoder (d) block of the proposed RE-DAE architecture as shown in eq. (5 & 6) and Fig. 2.

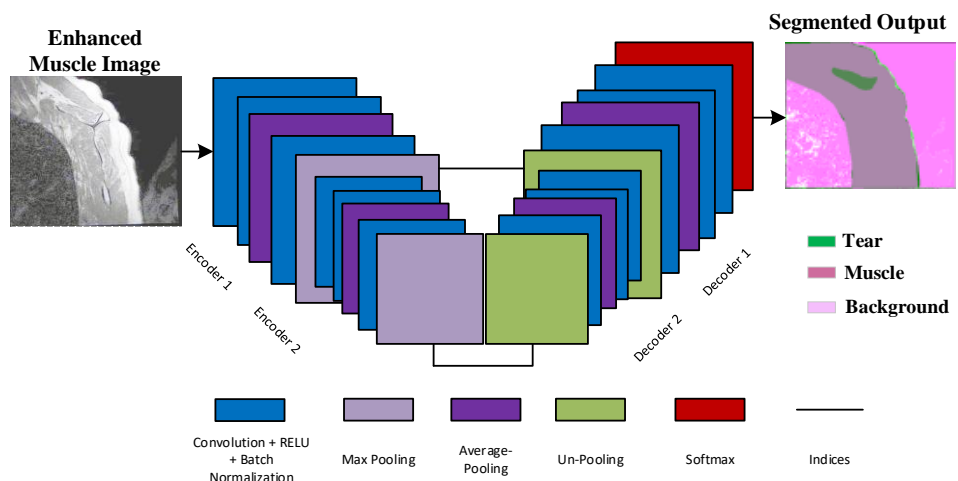

**Fig. 2.** The proposed (RE-DAE) CNN architecture uses two encoders and their corresponding matching decoders, with the final layer of pixel-based classification. The first encoder referred to as 'encoder1' is mapped with the first decoder named 'decoder1', and similarly, the second encoder mapped with the second decoder. Convolution layers, batch normalization, constitute the encoder network and activation function followed by $\mathbf{F}_{RE-e}$ operations as shown in eq. 5, and as a result, robust low-resolution hybrid (smooth and boundaries) feature sets are achieved. Similarly, in the decoder convolution network, first, we up-sample (Un-pool) the low-resolution feature sets and then, convolution layer followed by batch normalization is employed. Finally, the smoothing operation is applied to produce dense hybrid features ($\mathbf{F}_{RE-d}$) as illustrates in eq. 6, the same in spatial resolution to their corresponding input resolution.

### 2.2.2 Custom Deep CNN Segmentation Models

The existing CNN architectures are customized by modifying the initial and final layers to be compatible with the input feature map dimensions and output multi-class challenge (3-class). The following customized CNN models are employed for automatic muscle segmentation. Fig. 3 shows the details of customized models.

**Custom VGG (Visual Geometry Group)**

The VGG-16/19 encoder have 13/16 Convolutional and 3 fully connected (FC) layers [37]. These architectures were initially designed for object classification and achieved the second position in classification at the 2014-ILSVRC competition [38]. VGG, for the first time, replaced large filter size 11x11, 7x7, and 5x5 with 3x3 and experimentally verified that a small size filter makes the receptive field effective and produces excellent results as compared with large size filters. Secondly, the use of a small size filter reduces the number of parameters and reduces low computation complexity. VGG was initially designed for object recognition and classification. However, the classification layers (FC layers) are replaced with convolution layers at the decoder side to retain high-resolution and same dimension feature maps at the deepest corresponding input encoder for the segmentation task.

**Custom SegNet**

SegNet is a semantic pixel-wise segmentation CNN model [39]. SegNet has been initially proposed for outdoor scene understanding applications, specifically for an autonomous car, but recently has been effectively used for indoor scene understanding. SegNet has limited applications for indoor scenes, especially in the field of medical sciences. The segmentation architecture of SegNet, as shown in Fig. 3, is similar to VGG16 but has a different number of encoders and decoders. VGG has 5, while SegNet contains 2 encoders and their corresponding decoders [38]. Each encoder and decoder contains non-linear processing layers. Usually, at the end of each encoder network, non-linear down-sampling is performed using the max-pooling, and the high-frequency indices are stored. Each corresponding decoder network gets sparse low-resolution feature maps. Then, up-sampling is performed using the already stored indices calculated during the corresponding encoder's max-pooling operation. Also, the other advantages of pooling operations are: (1) retains the location information in terms of high-intensity of the segmented images, (2) Reduces the total number of trainable parameters during un-pooling at the decoder side.

**Custom FCN (Fully Convolution Network)**

The VGG-16 network is customized into FCN 32s, FCN 16s, and FCN 8s networks, having different up-sampling factors. FCN FC layers are replaced with fully convolution layers for object segmentation, and up-sampling is performed through transposed convolution layers. In FCN 32s at the decoder side, stride size 32 (non-overlapping window) was employed [40]. As a result, the resulting feature map was up-sampled by a factor of 32 across both width and height. FCN 32s extracts coarser features that are helpful for object classification, detection, and localization. On the other hand, FCN16s and FCN 8s extract both coarse and fine features and improve segmentation performance [41]. Both FCN16s and FCN 8s up-sample the low-level resolution contents obtained from the encoder network to get a precise output using 16 and 8 up-sampling factors, respectively.

In FCN 16s, the outcome of the network is the product of the up-sampling of two layers. These layers are up-sampled from the 4th pooling layer and (7th Convolutional layer) x 2. But in FCN 8s, the outcome is a product of the up-sampling of three layers. These layers are the up-sampling of the 3rd pooling layer, (4th pooling layer) x 2 and (7th Convolutional layer) x 4.

**Custom U-Net**

U-Net architecture is based on FCN and is considered one of the most popular convolution encoder-decoder architecture for biomedical applications [42], [43]. Comparing U-Net with FCN-8, there are two main differences: First, U-net is symmetric, and second, the skip connections are introduced between encoder and decoder feature map. The skip connection in segmentation architecture combines the same resolution features maps of both encoder and decoder in the up-sampling path to use both the coarse and fine features. The architecture is computationally efficient and trained end to end on a small amount of data. The encoder decreases the spatial dimension by using the down-sampling layer to extract the finely detailed features. While on the decoder side, these fine feature maps are up-sampled through transposed layers to get coarse detail for the segmentation (of the same dimensions to the corresponding input resolution map). U-Net can be directly be employed on 2D volume as well as on 3D volume data. While employed on the 3D volume data, it is computationally expensive and consumes high memory [44].

**Custom U-SegNet**

U-SegNet combines both SegNet and U-Net CNN architectures, largely used for semantic segmentation [45]. Most of the contemporary reported DL based semantic segmentation architectures are computationally expensive and require many learning parameters. These architectures also require a large amount of training dataset, but the availability of extensive medical data labeled by a pathologist is generally difficult. This motivates developing a U-SegNet architecture that gives a reasonable performance for medical image segmentation, provided that the amount of available data is limited and has a smaller number of parameters.

In U-SegNet, skip connection is incorporated in SegNet inspired from U-Net architecture to concatenate both coarser and more delicate details for final pixel-wise segmentation and reduce the number of parameters by using a 1x1 Convolutional layer. Because, at the decoder side, SegNet up-samples the coarser detail by using an un-pooling layer and utilizes a smaller number of parameters and trains quickly. Similarly, U-Net, by applying transposed Convolutional layer while up-sampling, starts from the extraction of coarser features and reaches to finer level information. As compared to SegNet, U-Net uses the concatenation of multi-scale information at the decoding side, which requires many learning parameters and trains slowly compared to SegNet. The U-SegNet architecture focuses on the local information as compared with global information for the segmentation. Because U-SegNet divides images into small patches, these patches are enough to achieve adequate information for segmentation. The depth of SegNet architecture decreases to the input patch size for compatibility. U-Net skips a connection at upper layers to incorporate feature maps with fine detail and reduces the number of parameters. U-SegNet at the decoding side uses a 1x1 Convolutional layer to concatenate both coarse and fine details for segmentation. It also reduces the number of parameters for the final convolution layer.

**Custom FCN-AlexNet**

FCN-AlexNet is a custom-made CNN architecture based on both the FCN and Alexnet network, as shown in Fig .3. AlexNet has been employed on ImageNet for object classification and won the ImageNet competition in 2012. Generally, Convolutional layers are used for feature extraction and maintain spatial information, while FC layers are used for feature classification. The network is made competitive for the segmentation task by replacing the FC layer of Alexnet with a corresponding Convolutional layer and up-sampling layers.

### 2.2.3 TL-based Fine-Tuned Segmentation Model

A CNN can effectively perform when the training data is abundant and has a 1D (vector) or 2D (image) grid-like topology. When enough amount of data is not available, the training performance of CNN mostly suffers. Especially in the medical field (muscle dataset), sometimes a small amount of labeled dataset is available. Therefore, TL has been utilized to produce a satisfactory performance on a limited amount of medical datasets [46]–[48]. TL is one of the most popular methods in ML that allows us to build an accurate network and save time and resources. Instead of training CNN from scratch, TL methods have been used to exploits the pre-trained network. The pre-trained networks were trained on different benchmark datasets, and suitable weights and biases have been obtained by solving a similar task to the problem-specific domain.

VGG16 and VGG19 CNNs were considered to have been already trained on the ImageNet dataset [7], [49]. The ImageNet dataset contains 15 million images of 1000 different classes, such as a mouse, keyboard, pencil, animals, etc. The network learned prominent features due to the considerable variation in the dataset. These pre-trained networks are imported and fine-tuned on problem-specific muscle datasets to exploit the concept of TL. The Fine-tuning of a pre-trained network with TL is much faster than training a network from scratch. The pre-trained network already learns the rich feature maps, and when fine-tuned, the network learns specific features according to the problem-specific domain. If the dataset is sufficient, then TL is not so fast as compared with learning from scratch. Similarly, SegNet [37], U-Net [38], Fully Convolutional Network (FCN) [50], named as FCN-8s, FCN-16s, and FCN-32s [40], were also initialized using layers and weights from the VGG network.

## 2.3 Pixel-label-based Segmentation

The CNN layers are used as a training network in the proposed RE-DAE, TL-based fine-tuned pre-trained, and custom-made semantic segmentation architectures. CNN layers in these architectures perform dynamic feature extraction in a stepwise fashion. The trained network's final layer is based on pixel classification in the last decoder, which performs pixel label-based segmentation. The softmax as an activation function was used to identify the class probabilities for each pixel. The pixel classification layer classifies each pixel based on the maximum feature maps and the MRI pixel values into three different regions, tear, muscle, and background.

## 2.4 Static Attention (SA)-based implementation of the Proposed RE-DAE

In this study, the idea of assigning static attention (SA) to each pixel of the dominant tear region (class) is incorporated in the proposed RE-DAE architecture [51], [52]. The proposed SA-RE-DAE enhances the tear region by assigning high weightage. In comparison, the background region is suppressed by assigning a less weight to their pixels. Weighted class name pairs perform the pixel classification, and cross-entropy is used as a loss function to achieve the final segmentation output. As a result, the semantic segmentation metrics and high visual quality of correct segmented regions are computed.

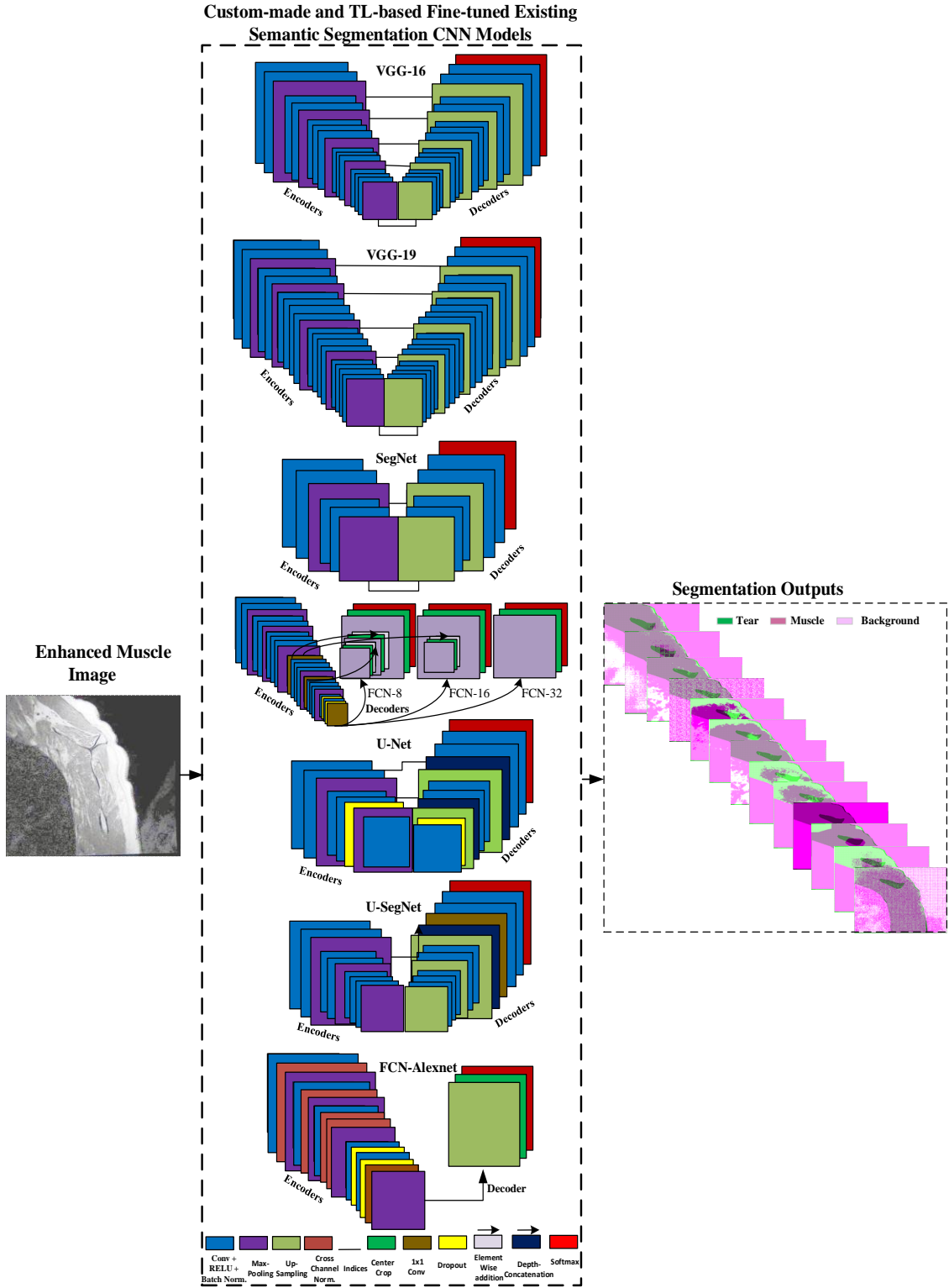

**Fig. 3.** Segmentation framework of the muscle MR images using customized CNN models

## 3. Experimental Setup

### 3.1  Dataset

The dataset was obtained from the Korea University Anam Hospital, Seoul, South Korea. The dataset contains 15 patients' images, and an augmentation strategy was employed to create more samples.

The dataset consists of mainly tear affected muscles. The ground truth was made through manual annotation as a pathological expert (the surgeon among the authors) provided the class knowledge. The tear affected images were partitioned into three classes named: tear, muscle, and background. Fig. 4 illustrates the pixel distribution of each region of the input image. Labels assigned to these classes were 0, 1, and 2 for background, muscle, and tear, respectively. All the images are in RGB format, and the foreground region is mostly located in the center. Images were captured from different views; therefore, the tear's size and position vary from different angles. Hence, the differences in the tear's size and position made the diagnosis of the tear very challenging. In practice, the expert physician knows the direction of captured MR images.

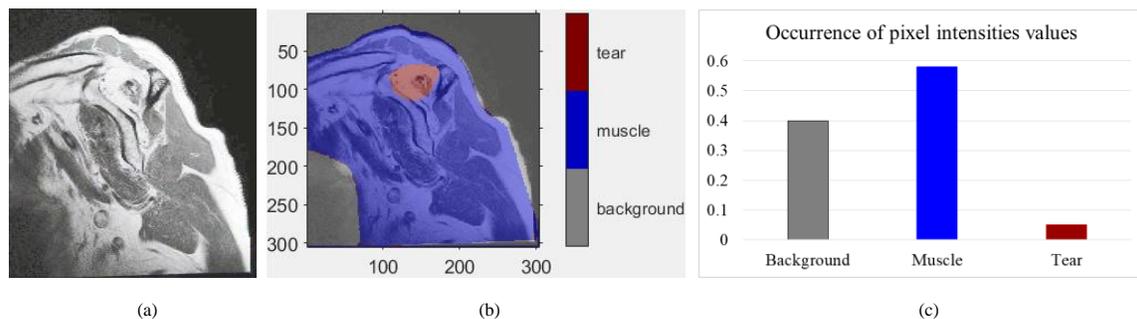

    (a)                              (b)                                    (c)
**Fig. 4.** (a) Shows pre-processed shoulder muscle image, (b) Color label, and (c) Pixel distribution of tear, supraspinatus muscle, and background regions, respectively.

### 3.2  Implementation Details

Regarding the network training, the hyper-parameter selection is important to control the optimization and convergence of the network. The "Stochastic Gradient Descent with Momentum (SGDM)" algorithm was used for optimization. The learning rate was kept $10^{-4}$, epochs = 30, the linear momentum = 0.9 and mini-batch = 2. Once the optimized parameters were selected, the validation dataset was also added to the training set.

All experiments were performed on MATLAB R2020b using Dell Core i7, 16 GB RAM. MATLAB DL toolbox was used for the implementation of segmentation. To train the CNN network is computationally expensive. Therefore, to reduce computational time, MatConvNet (a MATLAB based DL library) was used for experiments [53]. The total training time for each network took approximately 2 minutes on GPU NVIDIA GeForce GTX Titan X.

### 3.3  Performance Evaluation

Generally, the results of segmentation networks are compared by using standard measures like accuracy. But in the case of medical segmentation, accuracy is not enough to achieve the true picture of a network's performance. Therefore, accuracy, dice similarity (DS Score), and Jaccard Coefficient (intersection over union (IOU)) were used as performance metrics to evaluate the segmentation network's true performance. The evaluation of these parameters on each tested image was calculated using the manually annotated labels.

Mathematically, accuracy is expressed in eq. (7), which calculates the correct estimation against the total number of cases like global, local, and means accuracy. The accuracy has been considered for muscle segmentation since the number of accurate distributions of pixels to the specific class would affect the overall pixel classification result.

$$\text{Accuracy} = \frac{TP+TN}{TP+FP+FN+TN} \quad (7)$$

TP, FP is the amount of true and false-positive pixels. Whereas TN, FN is the amount of true and false-negative pixels of each pixel, respectively. Eqs. (8) & (9) mathematically express IOU and DS Score. For comparative analysis, the images are converted into label images as ground truth.

During testing, the pixel of each class was decided based on the majority voting of class labels obtained from overlapping patches. DS Score is an essential parameter for determining how the ground truth and simulated images are closely related. IOU is used to determine its close spatially matching with accurate mapping.

$$\text{Jaccard Coefficient (IOU)} = \frac{TP}{TP+FP+FN} \quad (8)$$

$$\text{Dice Similarity (DS Score)} = \frac{2*TP}{(2*TP+FP+FN)} \quad (9)$$

## 4. Result and Discussion
### 4.1 Results

Images of twelve patients from muscle datasets were considered for training (including augmented images), and the rest of them were used for testing. The highest performance was achieved when pathological data were augmented to train the proposed architectures SA-RE-DAE and RE-DAE. The result achieved with a DS score: 84.67%, 85.58% for the tear and 86.30%, 87.07% for the supraspinatus muscle region by employing SA-RE-DAE and RE-DAE on test 24 images, as shown in Table 1, respectively. The accuracy for the proposed segmented architectures RE-DAE and SA-RE-DAE was 81.57 %, 95.98 % and 80.52%, 95.26% for the tear and supraspinatus muscle region, respectively. The overall global accuracy and Weighted IoU were 85.73% and 74.27% for the proposed SA-RE-DAE. The cluster-wise accuracy and IoU are also mentioned in Table 1. The obtained results appear satisfactory and visually distinguishable. Fig. 5 (a) contains the original muscle MR image. Fig. 5 (b) shows a pre-processed image, while Fig. 5 (c) shows ground truth in RGB format. The segmented images using the proposed SA-RE-DAE and RE-DAE are also illustrated in Fig. 5 (d) & (e). In comparison, Fig. 5 from (f) to (u) shows the segmented image of the existing CNN models. The segmented image is represented by the color map and further compared with the labeled image to achieve the accuracy, DS, and IOU score of each class, namely; tear, and muscle. Furthermore, the segmented images using the proposed architectures are also illustrated in Fig. 6.

**Table 1**
Performance analysis of the proposed and Custom-made semantic segmentation models on test images.

| Model | Region | Objective Evaluation Metrics | | | | |
|---|---|---|---|---|---|---|
| | | DS % | Accuracy % | IOU % | Global Acc% | Weighed IOU% |
| **SA-RE-DAE** | Muscle | **87.07** | **95.98** | **77.43** | **85.73** | **74.27** |
| | Tear | **85.58** | **81.57** | **75.77** | | |
| **RE-DAE** | Muscle | **86.30** | **95.26** | **76.79** | **84.86** | **73.33** |
| | Tear | **84.67** | **80.52** | **74.81** | | |
| SegNet | Muscle | 84.30 | 93.13 | 75.22 | 83.21 | 71.29 |
| | Tear | 82.96 | 77.71 | 71.98 | | |
| FCN-8 | Muscle | 84.21 | 91.92 | 73.29 | 82.11 | 71.19 |
| | Tear | 82.77 | 77.22 | 72.12 | | |
| U-SegNet | Muscle | 84.05 | 94.19 | 73.48 | 81.82 | 68.92 |
| | Tear | 80.82 | 72.45 | 68.39 | | |
| UNet | Muscle | 83.55 | 93.12 | 75.69 | 81.97 | 70.82 |
| | Tear | 81.06 | 74.37 | 70.90 | | |
| VGG-19 | Muscle | 82.60 | 92.52 | 70.61 | 80.72 | 67.22 |
| | Tear | 78.43 | 75.18 | 68.47 | | |
| FCN-16 | Muscle | 81.45 | 94.18 | 71.13 | 79.51 | 66.16 |
| | Tear | 78.92 | 67.90 | 64.69 | | |
| VGG-16 | Muscle | 78.21 | 82.37 | 64.90 | 76.01 | 61.89 |
| | Tear | 76.53 | 74.26 | 62.88 | | |
| FCN-Alexnet | Muscle | 45.46 | 45.14 | 36.60 | 44.34 | 32.70 |
| | Tear | 40.17 | 43.73 | 34.24 | | |
| FCN-32 | Muscle | 41.46 | 35.94 | 26.60 | 34.76 | 24.70 |
| | Tear | 38.17 | 33.73 | 24.14 | | |
| Modified U-Net (UNet,VGG) [11]* | Muscle | 71.00 | --- | 72.71 | --- | --- |
| | --- | --- | --- | --- | | |

Modified U-Net [11]* results have been reported on OBPP muscle dataset [31].

## 4.2 Discussion

Initially, training the network has a high error rate; therefore, SGD fluctuates heavily to minimize the cross-entropy loss. With the decrease of loss, SGD movement becomes stable and converges to a better solution. After reaching convergence, the training was stopped. The obtained result shows that out of the entire validation set, the network could correctly classify pixels around 96% and 85 % of muscle and tear, respectively. The validation result shows that the segmentation network was fully trained on the problem-specific dataset. Additionally, the optimized parameters are obtained and now ready to perform segmentation on test images. Furthermore, the computed results of both custom and TL-based fine-tuned pre-trained semantic segmentation architectures have also been analyzed.

The maximum cluster accuracy within all the custom architectures against the maximum within all the TL-based fine-tuned pre-trained semantic segmentation architectures is (77.71 % against 80.99 %) for the tear region. Consequently, DS score and IOU are (82.96% against 83.52%) and (71.98% against 73.88%), respectively.

Accuracy scores for a segmented supraspinatus muscle and tear region are better. Supraspinatus muscle is thin, elongated in shape and variable contrast across the patients. Therefore, it recommends that muscles with very strong injury (tear) may be treated separately with weighted attention, only using pathological labeled augmented data.

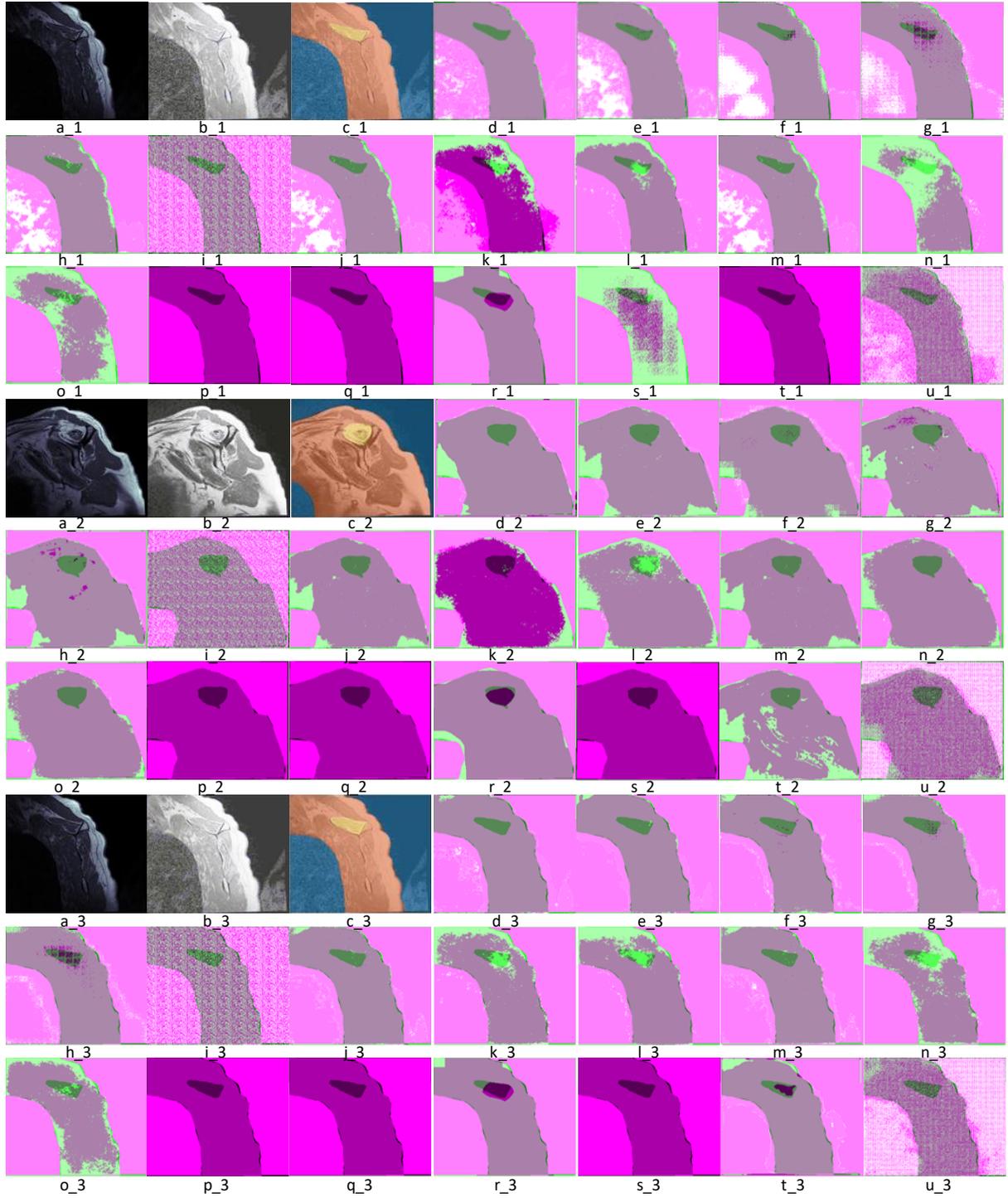

**Fig. 5.** (a) Shoulder muscle image, (b) Pre-processed image, (c) Color label, (d) the Proposed SA-RE-DAE, (e) the Proposed RE-DAE, (f) Pre-trained SegNet, (g) SegNet, (h) FCN-8, (i) FCN-16, (j) U-SegNet, (k) Pre-trained FCN-32, (l) FCN-Alexnet, (m) VGG19, (n) FCN-32, (o) Pre-trained FCN-Alexnet, (p) Pre-trained VGG16, (q) Pre-trained VGG19, (r) Pre-trained U-Net, (s) Pre-trained FCN-8, (t) VGG16, and (u) pre-trained FCN-16, shows segmented images. Whereas greenish, orchid, and magenta color represents tear, supraspinatus muscle, and background regions, respectively.

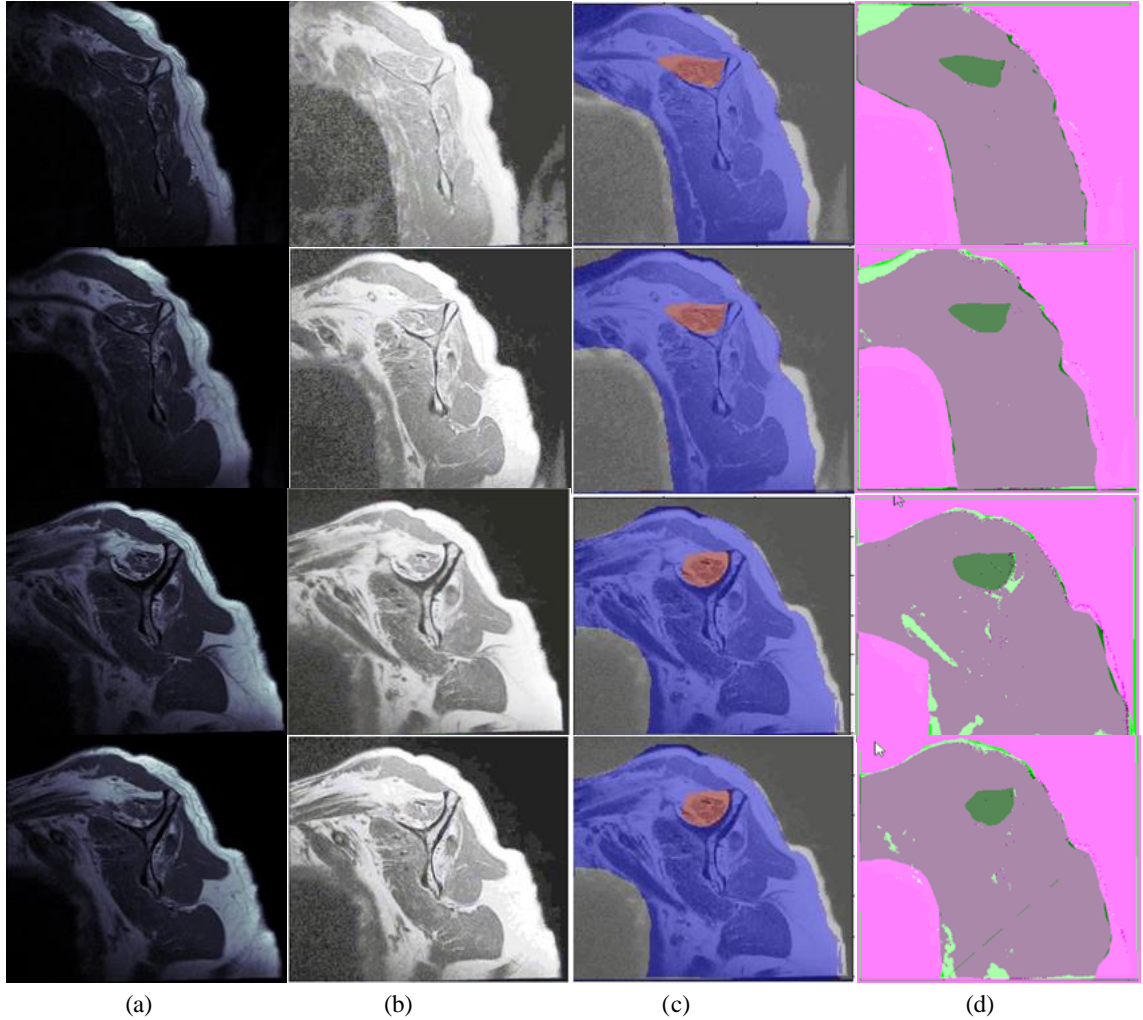

**Fig. 6.** (a) Shoulder Muscle Image, (b) Pre-Processed Image, (c) Color label, and (d) the Proposed SA-RE-DAE Segmented image. Whereas greenish, Orchid, and magenta color represents tear, supraspinatus muscle, and background regions, respectively.

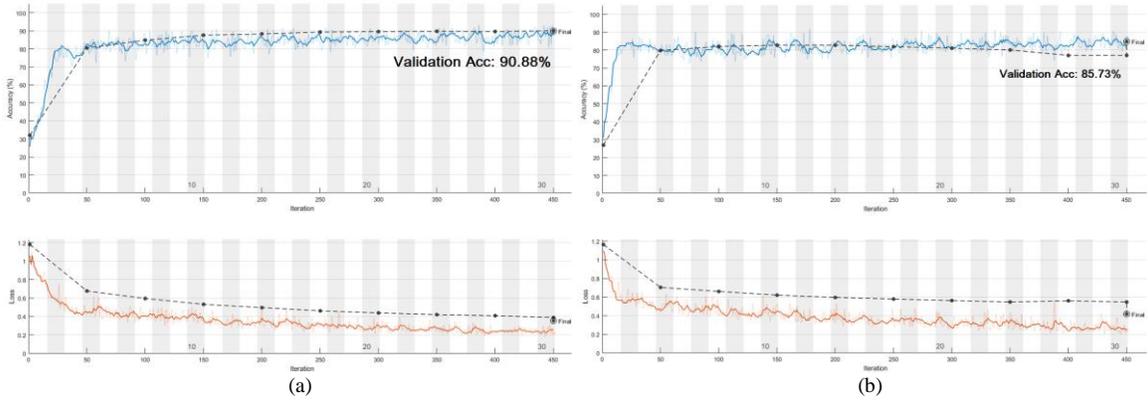

**Fig. 7.** Training and Validation plot of (a) RE-DAE, and (b) SegNet.

Moreover, the proposed SA-RE-DAE and RE-DAE appears to be globally suited from minor to major muscle impairments. Also, to improve the performance of the proposed architectures, the pathological and augmented data are combined and are considered to be a better strategy than TL-based fine-tuned pre-trained networks.

Furthermore, exploiting annotations for the diseased shoulder muscles, the same training strategy is also employed for custom, and TL-based fine-tuned semantic architectures.

## 4.3 Performance comparison of Proposed Architectures

We have validated the learning capacity of our proposed architectures RE-DAE and SA-RE-DAE by comparing their performance with different existing segmentation models. The proposed architecture globally outperforms existing custom, and TL-based fine-tuned the pre-trained semantic segmentation CNNs model. The detailed comparisons of all the models using the test data are illustrated in Table 1 & 2. The outperformance of the proposed architectures is due to incorporating a hybrid feature set, both smooth and boundary information of tear and muscle region for correct segmentation. Training accuracy and loss values for the proposed RE-DAE and the second-best model (Custom SegNet) are illustrated in Fig. 7. The plot clearly shows that RE-DAE has initially good convergence as compared to other architectures.

**Table 2**
Performance analysis of TL-Based fine-tuned pre-trained semantic segmentation models.

| Model | Region | Objective Evaluation Metrics | | | | |
|---|---|---|---|---|---|---|
| | | DS % | Accuracy % | IOU % | Global Acc% | Weighed IOU% |
| SegNet | Muscle | 84.63 | 88.93 | 74.91 | 82.84 | 71.32 |
| | Tear | 83.52 | 80.99 | 73.88 | | |
| FCN-8 | Muscle | 84.61 | 92.17 | 74.09 | 82.78 | 71.29 |
| | Tear | 83.77 | 77.90 | 72.40 | | |
| VGG16 | Muscle | 83.96 | 94.30 | 74.71 | 82.67 | 71.13 |
| | Tear | 81.19 | 75.18 | 71.82 | | |
| U-Net | Muscle | 83.95 | 92.12 | 74.39 | 82.52 | 71.07 |
| | Tear | 82.16 | 76.37 | 71.30 | | |
| VGG19 | Muscle | 83.21 | 92.82 | 70.92 | 81.12 | 68.21 |
| | Tear | 78.73 | 75.98 | 68.89 | | |
| FCN-16 | Muscle | 82.65 | 94.30 | 72.13 | 80.71 | 67.26 |
| | Tear | 79.29 | 67.90 | 65.49 | | |
| FCN-AlexNet | Muscle | 73.57 | 84.01 | 62.88 | 71.71 | 58.79 |
| | Tear | 70.43 | 62.21 | 57.66 | | |
| FCN-32 | Muscle | 61.56 | 61.14 | 46.60 | 64.76 | 45.70 |
| | Tear | 58.27 | 53.63 | 44.14 | | |

### 4.3.1 Segmentation using Static Attention (SA)-based RE-DAE

In the muscle data, the background and muscle regions dominate the tear region. This dominancy normally affects the segregation of tear region and as result reduces the performance of segmentation models. To address this issue, we incorporate a pixel attention strategy in the proposed RE-DAE. The pixel attention strategy improves the segmentation performance, evident from the subjective quality (Fig. 4 & 5) and objective performance measure (Fig. 8 & Table 1).

### 4.3.2 Performance comparison of TL-based fine-tuned and Custom-made models

The performances of TL-based CNN architectures are compared with custom-made. Each pre-trained CNN model has been trained on the different problem domains like ImageNet and then fine-tuned on the problem-specific muscle dataset using TL. The TL-based fine-tuned pre-trained architectures improved the performance from 5-7% in terms of accuracy compared to the corresponding custom CNN models. Whereas analyzing the DS Score, the DS performance is also increased to 5-6%, as shown in Fig. 8. Both the Custom and TL-based fine-tuned pre-trained segmentation CNN has the approximately same model. But the difference is: in custom-made, the initial and final layer is customized, and no learned patterns are transferred from a pre-trained architecture. The custom models collectively performed poorly, as shown in Fig. 8, as compared to the fine-tuned pre-trained CNN models. The poor performance of custom segmentation models shows that these models have difficulty in convergence due to the depth of CNN and the small size of the training set. Custom semantic models are trained from scratch on the muscle dataset. However, it is observed that these models like VGG, SegNet, and U-Net, etc., have previously shown good performance on the availability of a large amount of dataset for different domains like outdoor as well as indoor. This motivates that incorporating TL into these segmentation models might improve the segmentation performance. Fig. 8 and Table 1-2 clearly show that TL-based fine-tuning of the pre-trained architectures achieves a little improvement in segmentation results compared to just using custom semantic architectures. One reason for this slight improvement is the different source domains (natural outdoor or indoor dataset like ImageNet) and the target domains (medical dataset of the muscle). Individually, the worst-segmented model (FCN-32) DS result is 38.17% and 41.46% for the tear and muscle region, respectively. The poor performance of FCN-32 is due to up-sampling with a high rate (up=32) which may lose important information. Moreover, pre-processing methods also improved the accuracy of the proposed architecture and refined by an efficient data augmentation strategy.

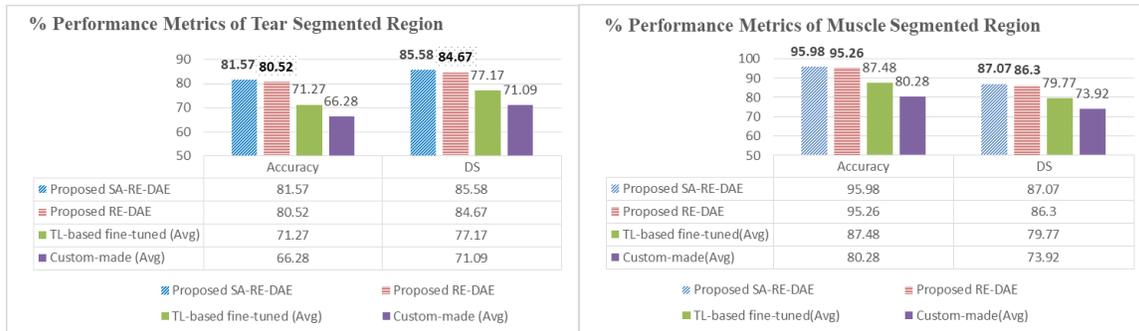

**Fig. 8.** Performance comparison of the proposed architectures (SA-RE-DAE, RE-DAE), TL-based fine-tuned (average) and custom-made (average) semantic segmentation models.

## 5. Conclusion and Future Work

This research study presents an automatic shoulder muscle segmentation using a new semantic segmentation RE-DAE architecture. The regional homogeneity and related boundary details are exploited by incorporating the concept of region and edge method, which is systematically employed in the encoder and decoder block of the proposed RE-DAE. This, in turn, enables the proposed technique to automatically segment closely related tear regions of shoulder muscle in MR images using pixel-label-based segmentation. Additionally, the pixel weightage (static attention idea) has been incorporated in the tear region. We employed different custom-made and TL-based fine-tuned semantic segmentation CNN architectures for comparative analysis. The proposed segmentation SA-RE-DAE model outperforms the existing segmentation models (DS:

85.58% and 87.07%, accuracy: 81.57% and 95.58 %, for tear and muscle region). In future, the proposed region and edge concept may be employed in the existing semantic segmentation models to increase the segmentation performance. The proposed architectures have the potential to be applied to other muscle types in applications such as neuromuscular diseases and other muscle diseases.

**Acknowledgment:**
This study is supported by the research grant of National Research Foundation of Korea (2017R1A2B2005065).This work is also conducted with the support of the PIEAS IT endowment fund under the Pakistan Higher Education Commission (HEC). As well as, we thank Pattern Recognition Lab (PR-Lab) and Pakistan Institute of Engineering, and Applied Sciences (PIEAS), for providing necessary computational resources and a healthy research environment.

**Conflict of Interest:**

Authors declared no conflict of interest.